\documentclass[12pt]{article}
\usepackage{graphicx}

\def\beqq{\begin{equation}}   
\def\eeqq{\end{equation}}
\def\bea{\begin{eqnarray}}   
\def\eea{\end{eqnarray}}

\newcommand{\GeV}{\,\mbox{GeV}}
\newcommand{\MeV}{\,\mbox{MeV}}
\newcommand{\matel}[3]{\langle #1|#2|#3\rangle}

\begin{document}
\begin{flushright} 
UND-HEP-04 -BIG\hspace*{.08em}10\\
hep-ph/0501084
\end{flushright}

\vspace*{4mm}

\centerline{\large\bf The Unreasonable Success of CKM Theory 
\footnote{Invited talk given at FPCP04 in Daegu (Korea), Oct. 5 - 10, 2004}}

\vskip 0.3cm \centerline{I. I. Bigi} 

\vskip 0.3cm 
\centerline{Dept. of Physics, University of Notre Dame du Lac, Notre Dame, IN 46556, U.S.A.}
\centerline{e-mail: ibigi@nd.edu}
\vskip 1.0cm

\centerline{\bf Abstract}
CKM theory has scored novel successes recently, 
namely vindication of weak universality and its first decisive test in CP studies. While this does not weaken the case for New Physics, one cannot count on the latter to induce 
large discrepancies in future studies. To establish smallish deviations one needs precise predictions; those in turn require accurate CKM parameters. I list strategies for extracting 
$V(cb)$, $V(ub)$ and $V(td)$. After short comments on New Physics  in semileptonic 
$B$ decays and on $\Delta \Gamma (B_s)$ I review the CP phenomenology in $B$ 
decays and stress that a comprehensive program of studying flavour dynamics has 
to form an essential complement of our next adventure, uncovering the physics behind the 
electroweak phase transition.

\section{CKM's Successes in the Last Millenium}
\label{INTRO}

The quark masses and CKM parameters control a plethora of observables in flavour dynamics;  
originating from nondiagonal quark mass matrixes, they presumably reflect dynamics at  
high, possibly even GUT scales. As long as flavour dynamics are described by a {\em single} 
$SU(2)$ group, the CKM matrix $V_{CKM}$ has to be unitary leading to the following constraints: 
$\sum _k V^*(ik)V(kj) = \delta_{ij}$. 
The three relations with $i=j$ describe `weak universality', the six with $i\neq j$ can be represented by triangles in the complex plane. There are four independent physical parameters: three angles and one complex phase. 
The latter is the sole source of CP violation (except for the `strong CP problem')   
and implies that all six triangles despite their vastly different shapes have the same area: 
${\rm area} = \frac{1}{2} J \; , \; J = |{\rm Im}[V^*(ub)V(ud)V(td)^*V(tb)]|$. 
Among these triangles there is one, which controls $B$ decays and is usually referred to as 
{\em the} CKM triangle (UT).  
Due to the `long' $B$ lifetime the lengths of its three sides are all 
${\cal O}({\rm sin}^3\theta _C)$ implying its three angles are `naturally' large. 

Since CKM theory with its few parameters undertakes to describe a huge body of data on flavour dynamics, it implies correlations between many a priori unrelated observables, in particular  
CP sensitive and insensitive quantities. The last point means that one can infer the size of the  
UT's angles, which control CP asymmetries, from the lengths of their sides, which have 
{\em no direct} sensitivity to CP violation. 
 \begin{figure}[h]
\centerline{
\includegraphics[height=10cm]{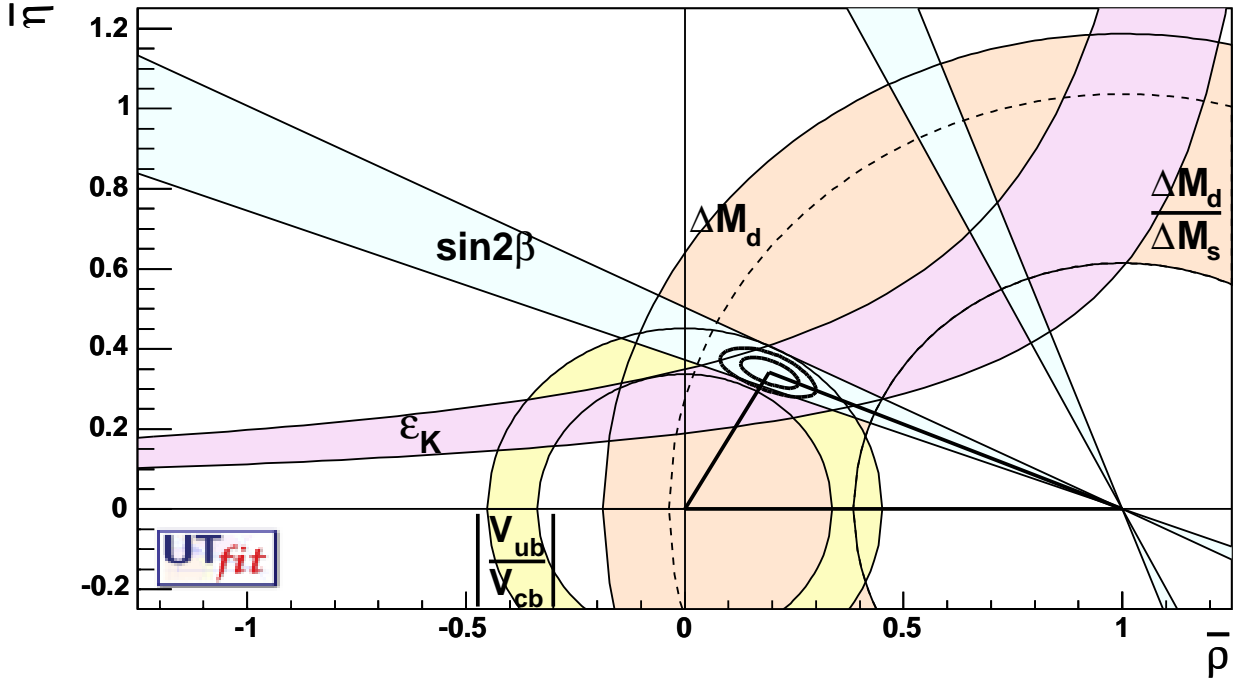}}

\caption{The allowed region for various observables  in the CKM fit.
\label{CKM04}}

\end{figure}

It is often stated that the data impose constraints on the UT that are represented by broad 
bands reflecting mainly theoretical uncertainties, see Fig.\ref{CKM04} \cite{UT}. While this is factually correct, it misses the bigger picture. The fact that the theoretical predictions for so diverse observables like $\Delta M_K$, $\Delta M_B$, $\epsilon_K$, $\epsilon^{\prime}$, which cover seven orders of magnitude when represented on an energy scale, come within 50 \% 
or so of the experimental values is most remarkable, 
in particular when one considers the input values $|V_{us}| \sim 0.22$, 
$|V_{cb}|, |V_{ts}| \sim 0.04$, $|V_{ub}|, |V_{td}| \sim 0.004$, $m_u \sim 5 \MeV$, 
$m_d \sim 10 \MeV$, $m_s \sim 100 \MeV$, $m_c \sim 1.2 \GeV$, $m_b \sim 4.6 \GeV$, 
$m_t \sim 180 \GeV$. Those values 
would have seemed frivolous choices -- if they had not been enforced upon us by other observables. 
There could easily have been massive inconsistencies even before the CP asymmetry in 
$B_d \to \psi K_S$ had been measured. While this asymmetry, expressed through sin$2\phi_1$, had been predicted to be at least sizeable \cite{BS80}, courageous people predicted in 1998 based on the CKM phenomenology known at that time
\beqq 
\left. {\rm sin}2\phi_1\right|_{'98\, pred.} = 0.72 \pm 0.07 \; . 
\label{PRED98} 
\eeqq

\section{CKM in the New Millenium}
\label{CKMNEWMILL}

BELLE and BABAR analyses of $B_d(t) \to \psi K_S$ are in impressive agreement  
\beqq
\left. {\rm sin}2\phi_1\right| _{WA '04} = 0.726 \pm 0.037 \; ,  
\eeqq
consistent with Eq.(\ref{PRED98}) as well as the '04 CKM phenomenology, Fig.\ref{CKM04}. The situation can be characterized as follows: 
{\bf (1)} 
The CKM paradigm has been promoted to a {\em tested theory}. 
{\bf (2)} 
CP violation has been `demystified': if the dynamics are sufficiently complex for supporting  
CP violation, the latter does not have to be small, i.e. the observable phases can be large. 
{\bf (3)}  
This `demystification' will be completed, once one finds CP violation anywhere in the lepton sector. 
{\bf (4)}  
CKM theory relates the `near miss' of CP invariance in $K_L$ decays naturally to the fact that 
the first and second quark families are almost decoupled from the third (although it does not explain 
that fact itself). 
{\bf (5)} 
Standard CKM dynamics are irrelevant for the baryon number of the Universe. 

A look at the present constraints, Fig.\ref{CKM04} shows that without $\epsilon_K$ and 
sin$2\phi_1$ the CKM triangle could still be completely flat. Yet if $\Delta M_{B_s}$ 
is found not far above the present lower bound the CKM description would have scored another 
impressive success. For we could conclude that the CP insensitive observables 
$|V_{ub}/V_{cb}|$ and $\Delta M_B$ imply CP violation  as observed! 

The CKM framework with its family structure and the very peculiar pattern of its parameters cries out for an explanation embedding it into a more complete theory, and one can expect deviations from its 
predictions, yet one cannot count on them being large, let alone massive. I.e., a shift of a CP asymmetry by 20\% might well be on the large side of what can be expected realistically (apart for special cases 
like $B_s(t) \to \psi \phi$ \cite{BS80}). This skeptical expectation is based on the aforementioned CKM successes ranging over several orders of magnitude. It is unlikely that a generic type of New Physics could turn the same remarkable trick (unless it is some version of SUSY). That means that {\em precision -- on the experimental as well as 
theoretical side -- has to be the order of the day.} 

\subsection{The Sides of `the' Triangle} 
\label{SIDES}

The status concerning heavy quark parameters (HQP) achieved can be summarized as follows:
\bea 
m_b^{kin} (1 \; \GeV) &=& 4.61 \GeV \pm 1.5\% , \, m_c^{kin} (1\; \GeV) = 1.18 \GeV \pm 7.8\%   \\
|V(cb)| &=& 41.390 \cdot 10^{-3} \pm 2.1\% \;  ;  
\label{STATUS}
\eea
i.e., a precision quite comparable to the `gold standard' of strange physics: 
\beqq 
|V(us)| = 0.2248 \pm  1\% \; \; \; {\rm Ref.} \cite{MESCIA}, 
\;  \bar m_s (2\, \GeV) =  78 \MeV \pm 13 \%   \; \; \; {\rm Ref.} \cite{HASHIM} 
\eeqq
The tale behind these numbers form a paradigm: a {\em robust} theoretical 
framework subjected to the challenges of high quality data can produce small uncertainties that can be {\em defended}. 

\subsubsection{Weak Universality and the Fall \& Rise of V(us)}
\label{VUS}

For the weak universality relation 
\beqq 
|V(ud)|^2 + |V(us)|^2 +|V(ub)|^2 = 1
\label{WU1}
\eeqq
only the first two 
terms are significant. After considerable scrutiny of the determination of $|V(ud)|$, the attention has focussed on $V(us)$. 
Eq.(\ref{WU1}) implies: 
\beqq
|V(us)|^{unit} = 0.2265 \pm 0.0022
\eeqq
Previously accepted values of $|V(us)|$ from $K_{l3}$ decays implied a 2.2 $\sigma$ 
deviation. This discrepancy has disappeared due to new data  
by KTeV, KLOE \& E865 \cite{SASHA}, from which one infers: 
\beqq
|V(us)|_{Kl_3} = \left\{ 
\begin{array}{ll}  0.2248 \pm 0.0005|_{exp} \pm 0.0018|_{theor} & {\rm Ref.\cite{MESCIA}}\\
0.2250 \pm 0.0010|_{exp} \pm 0.0020|_{theor} & {\rm Ref.\cite{OKA}}
\end{array}
\right. \;  ; 
\label{VUS}
\eeqq
The leading source for the theoretical uncertainty comes from the $SU(2)$ breaking of the 
 form factor $f_+(0)$ and its dependance on kinematical variables beyond the linear ansatz. 

\subsubsection{The $V(cb)$ Saga}
\label{VCB}

{\bf Inclusive Rates:} The value of $|V(cb)|$ is extracted from $B\to l \nu X_c$ in two steps. 

{\bf A:} 
One expresses $\Gamma (B \to l \nu X_c)$ in terms of the HQP -- quark masses 
$m_b$, $m_c$ and the expectation values of local operators $\mu_{\pi}^2$, $\mu_G^2$, 
$\rho_D^3$ and $\rho_{LS}^3$ -- as accurately as possible, namely through 
${\cal O}(1/m_Q^3)$ and to all orders in the BLM treatment for the partonic contribution. 
Those HQP are actually of two types, namely 
$m_{b,c}$ which are {\em external} to QCD and the expectation values that are 
{\em internal} to QCD. Having 
precise values for these HQP is not only of obvious use for extracting $|V(cb)|$ and $|V(ub)|$, 
but also yields benchmarks for how much numerical control lattice QCD provides us over 
nonperturbative dynamics. 

{\bf B:}  
The numerical values of these HQP are extracted from the {\em shapes} of inclusive 
lepton distributions as encoded in their {\em normalized} moments.  Two types of moments have 
been utilized, namely lepton energy and hadronic mass moments. While the former are dominated by the contribution from `partonic' term $\propto \matel{B}{\bar bb}{B}$, the latter are more sensitive to higher nonperturbative terms $\mu_{\pi}^2$ \& $\mu_G^2$ and thus have to form an integral part of the analysis. 
  
Executing the first step in the so-called kinetic scheme and inserting the experimental number for 
$\Gamma (B\to l \nu X_c)$ one arrives at \cite{BENSON1} 
\bea
\nonumber
\frac{|V(cb)|}{0.0417} &=& D_{exp}\cdot (1+\delta _{th})  [1+0.3 (\alpha_S(m_b) - 0.22)]  
\left[ 1 - 0.66(m_b - 4.6) + 0.39(m_c - 1.15)  \right. \\ 
\nonumber 
&& \left. +  0.013(\mu_{\pi}^2 - 0.4) + 0.05 (\mu_G^2 - 0.35) + 0.09 (\rho_D^3 - 0.2) + 
0.01 (\rho_{LS}^3 + 0.15 )\right]  \; , \\
&& D_{exp} = \sqrt{\frac{\rm BR_{SL}(B)}{0.105}}\sqrt{\frac{1.55\, {\rm ps}}{\tau_B}}
\label{VCBHQP}
\eea
where all the HQP are taken at the scale 1 GeV and their `seed' values are given in the 
appropriate power of GeV; the theoretical error at this point is  given by 
\beqq 
\delta _{th} = \pm 0.5 \%|_{pert} \pm 1.2 \% |_{hWc} \pm 0.4 \% |_{hpc} \pm 0.7 \% |_{IC} 
\eeqq 
reflecting the remaining uncertainty in the Wilson coefficient of the leading operator 
$\bar bb$, as yet uncalculated  perturbative corrections to the Wilson coefficients of the 
chromomagnetic and Darwin operators, higher order power corrections including 
duality violations in $\Gamma _{SL}(B)$ and nonperturbative effects due to operators containing 
charm fields, respectively.  

BaBar has performed the state-of-the-art analysis of several lepton energy 
and hadronic mass moments \cite{BABARVCB} obtaining 
an impressive fit with the following HQP in the kinetic scheme \cite{SCHEMES}: 
\bea 
m_b(1 \, \GeV)  = (4.61 \pm 0.068) \GeV , \, m_c(1 \, \GeV) = (1.18 \pm 0.092) \GeV  
\label{MB}\\
m_b(1 \, \GeV) - m_c(1 \, \GeV) = (3.436 \pm 0.032) \GeV  
\label{MBMMC}\\ 
\mu_{\pi}^2 (1\, \GeV) = (0.447 \pm 0.053) \GeV ^2 , \, 
\mu_{G}^2 (1\, \GeV) = (0.267 \pm 0.067) \GeV ^2 
\label{MUPI} \\
\rho_{D}^3 (1\, \GeV) = (0.195 \pm 0.029) \GeV ^3 
\label{RHOD}\\
|V(cb)|_{incl} = 41.390 \cdot (1 \pm 0.021) \times 10^{-3} 
\label{VCBBABAR}
\eea
The DELPHI collab. has refined their pioneering study of 2002 obtaining \cite{DELPHI}: 
\beqq
|V(cb)|_{incl} = 42.1 \cdot (1 \pm 0.025) \times 10^{-3}  \; , {\rm preliminary}
\label{DELPHI}
\eeqq

For a full appreciation of these results one has to note the following: 
\begin{itemize}
\item 
One obtains an excellent fit also when one `seeds' two of these HQP to their predicted values, namely 
$\mu_G^2 (1\, \GeV) = 0.35 \pm 0.03 \GeV^2$ as inferred from the $B^*-B$ hyperfine mass splitting 
and $\rho_{LS}^3 = - 0.1 \GeV^3$ allowing only the other four HQP to float. 
\item 
While these HQP are treated as free fitting parameters, they take on very special values fully consistent with all constraints that can be placed on them by theoretical means as well as other experimental input. To cite but a few examples: the rigorous inequality $\mu_{\pi}^2 > \mu_G^2$ is satisfied, 
$\mu_G^2$ indeed emerges with the correct value, see Eq.(\ref{MUPI}); the value for 
$m_b$ inferred from the {\em weak} decay of a $B$ meson, Eq.(\ref{MB}), agrees completely within the 
stated uncertainties with what has been derived from the {\em electromagnetic} 
and {\em strong} production of $b$ hadrons just above threshold as first suggested by Voloshin 
\cite{URIREV}: 
\beqq 
m_b (1\, \GeV)|_{e^+e^- \to \bar bb} = 4.57 \pm 0.08 \GeV \; ; 
\eeqq
$m_b$-$m_c$, Eq.(\ref{MBMMC}) agrees very well with what one infers from $B$ and $D$ meson masses \cite{URIREV}: 
\beqq
m_b(1\, \GeV) - m_c (1\, \GeV)|_{M_B-M_D} \simeq 3.48 \pm 0.02 \GeV
\eeqq
However this {\em a posteriori} agreement does {\em not} justify imposing it as an {\em a priori}  constraint. 
\item 
The $1.5$ \% error in $m_b$ stated in Eq.(\ref{MB}) taken at face value might suggest that it alone would generate more than a 3.5 \% uncertainty in $|V(cb)|$, i.e. considerably larger than 
the error given in Eq.(\ref{VCBBABAR}). The resolution of this apparent contradiction is as follows.  
The dependance of the total semileptonic width and also of the lowest lepton energy 
moments on $m_b$ \& $m_c$ can be approximated 
by $m_b^2(m_b - m_c)^3$ for the actual quark masses; for the leading contribution this can 
be written as $\Gamma _{SL}(B) \propto (m_b - \frac{2}{3}m_c)^5$. From the values for 
$m_b$ and $m_c$, Eq.(\ref{MB}), and their correlation given in \cite{BABARVCB} one derives 
\beqq 
m_b(1 \, \GeV) - 0.67 m_c(1 \, \GeV) = (3.819 \pm 0.017) \GeV = 3.819\cdot (1\pm 0.45\%)\GeV . 
\label{MBM23MC}
\eeqq
It induces an uncertainty of 1.1 \% into the value for $|V(cb)|$. 

\end{itemize}
With all these cross checks we can defend the smallness of the stated uncertainties. The analysis of Ref.\cite{GLOBAL} arrives at similar 
numbers (although I cannot quite follow their error analysis). 

More work remains to be done: (i) It is important for BELLE to perform a similarly 
comprehensive analysis. (ii) The errors on the hadronic mass moments are still 
sizable; decreasing them will have a significant impact on the accuracy of $m_b$ and $\mu_{\pi}^2$. 
(iii) As discussed in more detail below, imposing high cuts on the lepton energy degrades the 
reliability of the theoretical description. Yet even so it would be instructive to analyze 
at which cut theory and data part ways. (iv) As another preparation for $V(ub)$ extractions one 
can measure $q^2$ moments or mass moments with a $q^2$ cut to see how well one can 
reproduce the known $V(cb)$. 

\noindent{\bf Exclusive Modes:}
While it is my judgment that the most precise value for $|V(cb)|$ can be extracted from 
$B\to l \nu X_c$, this does not mean that there is no motivation for analyzing exclusive modes. On the contrary: the fact that one extracts a value for $|V(cb)|$ from $B\to l \nu D^*$ at zero recoil fully consistent within a smallish uncertainty represents a great success since the systematics experimentally as well as theoretically are very different: 
\beqq 
|V(cb)|_{B\to D^*} =  0.0416 \cdot (1\pm 0.022|_{exp} \pm 0.06|_{th}  )  
  \; \; \; \; {\rm for} \; \; \; \; F_{B\to D^*}(0) = 0.90 \pm 0.05
\eeqq
It has been suggested \cite{BPS} to treat $B\to l \nu D$ with the 
`BPS expansion' based on $\mu_{\pi}^2 \simeq \mu_G^2$ and extract 
$|V(cb)|$ with a theoretical error not larger than $\sim 2\%$. It would be most instructive to compare the formfactors and their slopes found in this approach with those of LQCD \cite{OKA}.

\subsubsection{The Adventure Continues: $V(ub)$}
\label{VUBSECT}

There are several lessons we can derive from the $V(cb)$ saga: 
(i) Measuring various moments of $B\to l \nu X_u$ and extracting HQP from them is a 
powerful tool to strengthen confidence in the analysis. Yet it 
is done for validation purposes only. For there is no need to `reinvent the wheel':
{\em When calculating the width and (low) moments of 
$B\to l \nu X_u$ one has to use the values of the HQP as determined in 
$B\to l \nu X_c$}. 
(ii) $\Gamma (B \to l \nu X_u)$ is actually under better theoretical control than 
$\Gamma (B \to l \nu X_c)$ since the expansion parameter is smaller -- 
$\frac{\mu_{had}}{m_b}$ vs. $\frac{\mu_{had}}{m_b-m_c}$ -- and 
${\cal O}(\alpha_S^2)$ corrections are known exactly. 

\noindent {\bf On the Impact of Cuts:} 
In  practice there arises a formidable complication: to distinguish 
$b\to u$ from the huge $b\to c$ background, one applies  
cuts on variables like lepton energy $E_l$, 
hadronic mass $M_X$, the lepton-pair invariant mass $q^2$. As a general rule the 
more severe the cut, the less reliable the theoretical calculation becomes. 
More specifically 
the imposition of a cut introduces a new dimensional scale called `hardness' {\cal Q}.  
Nonperturbative contributions emerge scaled by powers of $1/{\cal Q}$ rather than 
$1/m_b$. If {\cal Q} is much smaller than $m_b$ such an expansion becomes unreliable. 
Furthermore the OPE cannot capture terms of the form 
$e^{-{\cal Q}/\mu}$. While these are irrelevant for ${\cal Q} \sim m_b$, they quickly gain relevance 
when {\cal Q} approaches $\mu$. Ignoring this effect would lead to a `bias', i.e. a {\em systematic} 
shift of the HQP away from their true values. 

This impact has been studied for radiative $B$ decays with their simpler kinematics in a pilot study \cite{MISUSE} and a detailed analysis \cite{BENSON2} 
of the average photon energy  and its variance. The first provides a measure mainly of 
$m_b/2$, the latter of $\mu_{\pi}^2/12$. These biases were found to be relevant down to 
$E_{\rm cut} = 1.85\, \GeV$ and grow quickly out of control above 2.1 GeV. Data from CLEO and BELLE 
show some evidence for these predicted biases, in particular for the variance. Analyzing the 
energy moments as functions of the lower cut probes the lower tail of the spectrum. If the predictions for the moments as a function of the cut match the data over some range, one has 
demonstrated theoretical control in a new domain. Even more important than providing us with possibly more accurate values for $m_b$ and $\mu_{\pi}^2$, it would enhance confidence in our theoretical 
tools. Therefore I make the personal plea to BELLE and BABAR to present their findings on the photon energy moments for $E_{\rm cut}= 1.8$, $1.9$, $2.0$, $2.1$ and even $2.2$ GeV in the $B$ rest frame. 

These findings lead to the following conclusions: (i) As far as theory is concerned there is a high premium on keeping the cuts as low as possible. (ii) Such cuts introduce biases in the HQP values extracted from the truncated moments; yet within a certain range of the cut variables those biases can be corrected for and thus should not be used to justify inflating the theoretical uncertainties. 
(iii) In any case measuring the moments as functions of the cuts provides powerful cross checks for our theoretical control. 

\noindent{\bf `Let a Thousand Blossoms Bloom':} 
Several suggestions have been made for cuts to suppress the $b\to c$ background to 
managable proportions. None provides a panacea. 
The most straightforward one is to focus on the lepton energy endpoint region; however it captures 
merely a small fraction of the total $b\to u$ rate, which can be estimated only with considerable 
model dependance. This model sensitivity can be moderated with information on the heavy quark 
distribution function inferred from $B\to \gamma X$. Furthermore 
weak annihilation contributes only in the endpoint region and with different weight in $B_d$ and 
$B_u$ decays \cite{WA}. Thus the lepton spectra have to be measured {\em separately} for charged and neutral $B$ decays. 

Measuring the hadronic recoil mass spectrum up to a maximal value 
$M_X^{\rm max}$ captures the lion share of the $b\to u$ rate if $M_X^{\rm max}$ is above 1.5 GeV; yet it is still vulnerable to theoretical uncertainties in the very low $q^2$ region. This problem can be 
addressed in two different ways: adopting Alexander the Great's treatment of the Gordian knot 
one can  
impose a lower cut on $q^2$ or one can describe the low $q^2$ region with  the help of the measured  
energy spectrum in $B\to \gamma X$ for 1.8 GeV $\leq E_{\gamma} \leq$ 2.0 GeV. Alternatively 
one can apply a combination of cuts. Studying $B_d$ and $B_u$ decays is still desirable, yet not as 
essential as for the previous case. 

In any case one should not restrict oneself to a fixed cut, but vary the latter over some reasonable range 
and analyze to which degree theory can reproduce this cut dependence to demonstrate 
control over the uncertainties. 

There is not a single `catholic' path to the promised land of a precise value for 
$|V(ub)|$; presumably many paths will have to be combined 
\cite{BAUERPUERTO}. Yet it seems quite realistic that the 
error can be reduced to about 5 \% over the next few years.

\subsubsection{$V(td)$ from inclusive radiative $B$ Decays}
\label{VTD}

The usually quoted observables from which to extract $|V(td)|$ are: 
(i) $\Gamma (K^+ \to \pi ^+ \nu \bar \nu)$ -- theoretically the cleanest, yet experimentally the 
most challenging one; (ii) $\Delta M(B_d)/\Delta M(B_s)$ -- once  $\Delta M(B_s)$ has 
been resolved and an accurate value obtained for $B_{B_d}f^2_{B_d}/B_{B_s}f^2_{B_s}$; 
(iii) $\Gamma (B\to \rho \gamma )/\Gamma (B\to K^* \gamma)$ -- if one can obtain a reliable 
estimate for the hadronic form factors. 

With each of these strategies having its drawbacks, it would seem reasonable to study whether 
$B \to \gamma X_d$ can be distinguished against the dominant $B\to \gamma X_s$ 
at least over some major fraction of the 
spectrum with significant -- say 20\% --  accuracy. 

All these observables are complementary also in the sense that they exhibit different sensitivities 
to various New Physics scenarios.

\subsection{New Physics in Semileptonic $B$ Decays}
\label{NPSL}

Semileptonic $B$ decays are usually assumed to be insensitive to New Physics 
thus yielding the true values of the CKM parameters. Yet "trust is good -- control is better". 

We know that $b\to l \nu c$ is driven predominantly by SM $(V-A)_q\times (V-A)_l$ currents; the latest piece 
of evidence for it is the successful description of the measured lepton energy and hadronic mass moments in terms of a handful of HQP. It would be worthwhile to ask 
how much $(V+A)_q\times (V-A)_l$ admixture is allowed by the data on the moments 
\footnote{A $(V\pm A)_q\times (V+A)_l$ component would require a light right handed neutrino.}. 

The SM prediction for $\Gamma (B\to \tau \nu X_c)$ \cite{NIR} should be updated; for this width could be affected appreciably by the exchange of a charged Higgs with a mass in the 
several hundred GeV range for large tan$\beta$ scenarios. The exchange of such a Higgs can significantly affect also the exclusive channel $B \to \tau \nu D$ \cite{JAPANTAU}. 
However the hadronic form factors do {\em not} drop out of the ratio 
$\Gamma (B\to \tau \nu D)/\Gamma (B \to \mu \nu D)$. If the `BPS' approximation sketched above is 
validated by extracting the correct value of $|V(cb)|$ from $B\to e/\mu \nu D$, then it can be employed 
for calculating the two hadronic form factors controlling $B\to \tau \nu D$ on the few percent accuracy 
level.

\subsection{`CDF's June Surprise: a Large $\Delta \Gamma (B_s)$}
\label{CDF}

In June 2004 the CDF collaboration first presented an intriguing analysis exhibiting two 
surprisingly large lifetimes controlling $B_s \to \psi \phi|_{CP =\pm}$ \cite{CDF}: 
\bea
\tau [CP = +] = (1.05 ^{+0.16}_{-0.13} \pm 0.02) ps \; \; \; \; vs. \; \; \; \; 
\tau [CP = -] = (2.07 ^{+0.58}_{-0.46} \pm 0.03) ps \\
\frac{\Delta \Gamma _s}{\bar \Gamma_s} \equiv 
\frac{\Gamma (B_s[CP=+]) - \Gamma (B_s[CP=-])}
{\frac{1}{2}(\Gamma (B_s[CP=+]) + \Gamma (B_s[CP=-]) )}
 = (65^{+25}_{-33} \pm 1)\% 
 \label{CDFDELTA}
\eea
$\frac{\Delta \Gamma _s}{\bar \Gamma_s}$ being as large 
as 0.4 or even larger would open up a whole new realm of CP studies in 
$B_s$ decays with a great potential to identify New Physics; yet it would point at a severe 
limitation in our theoretical understanding of $B$ lifetimes. For the only channels that can 
produce an appreciable lifetime difference between the two $B_s$ mass eigenstates 
are driven by $b\to c \bar c s$. Let $R(b\to c \bar cs)$ denote their fraction of all $B_s$ decays. 
If these transitions contribute only 
to $\Gamma (B_s(CP=+))$ one has $\Delta \Gamma_s/\bar \Gamma_s = 2R(b\to c \bar cs)$. 
Of course this is actually an upper bound quite unlikely to be even remotely saturated. 
With the estimate $R(b\to c \bar cs) \simeq 25\% $, which is consistent with the data on 
the charm content of $B_{u,d}$ decays this upper bound reads 50\%. More realistic calculations 
have yielded considerably smaller predictions: 
\beqq 
\frac{\Delta \Gamma_s}{\bar \Gamma_s} = \left\{
\begin{array}{ll} 22\% \cdot \left(\frac{f(B_s)}{220\, \MeV}\right)^2 & {\rm Ref.\cite{AZIMOV}}\\
12 \pm 5\% & {\rm Ref.\cite{LENZ}}
\end{array}
\right. \;  ; 
\label{DELTAPRED}
\eeqq 
A value as high as $20 - 25$ \% is thus not out of the question theoretically, and Eq.(\ref{CDFDELTA}) is still consistent with it. One should note that invoking New Physics would actually `backfire' since it leads to a lower prediction.
If, however, a value exceeding 25\% were established experimentally, we had to draw at least one of the following conclusions: (i) $R(b\to c \bar cs)$ actually exceeds the estimate of 25\% significantly. This would imply 
at the very least that the charm content is higher in $B_s$ than $B_{u,d}$ decays by a 
commensurate amount and the $B_s$ semileptonic branching ratio lower. 
(ii) Such an enhancement of  $R(b\to c \bar cs)$ would presumably -- though not necessarily -- imply 
that the average $B_s$ width exceeds the $B_d$ width by more than the predicted 1-2\% level. 
That means in analyzing $B_s$ lifetimes one should allow $\bar \tau (B_s)$ to float 
{\em freely}. (iii) If in the end one found the charm content of $B_s$ and $B$ decays to be quite 
similar and 
$\bar \tau (B_s)$ $\simeq$ $\tau (B_d)$, yet  
$\Delta \Gamma_s/\bar \Gamma_s$ to exceed 0.25, we had to concede a loss of theoretical control 
over $\Delta \Gamma$. This would be disappointing, yet not inconceivable: the a priori reasonable 
ansatz of evaluating both  
$\Delta \Gamma_B$ and $\Delta M_B$ from quark box diagrams -- with the only 
manifest difference being that the internal quarks are charm in the former and top in the 
latter case -- obscures the fact that the dynamical situation is actually different. In the latter case the 
effective transition operator is a local one involving a considerable amount of averaging over 
off-shell transitions; the former is shaped by on-shell channels with a relatively small amount 
of phase space: for the $B_s$ resides barely 1.5 GeV above the $D_s \bar D_s$ threshold. 
To say it differently: the observable $\Delta \Gamma_s$ is more vulnerable to limitations 
of quark-hadron duality than $\Delta M_s$ and even beauty lifetimes 
\footnote{These are all dominated by nonleptonic transitions, where duality violations can be 
significantly larger than for semileptonic modes.} . 

In summary: establishing $\Delta \Gamma_s \neq 0$ amounts to important qualitative progress in our 
knowledge of beauty hadrons; it can be of great practical help in providing us with novel probes of CP 
violations in $B_s$ decays, and it can provide us theorists with a reality check concerning the reliability 
of our theoretical tools for nonleptonic $B$ decays.

\section{Angles of the Unitarity Triangles}
\label{ANGLES}

There are four classes of CP violating observables: 
(1) CP asymmetries involving $B^0 - \bar B^0$ oscillations; 
(2) direct CP violation; 
(3) CP violation {\em in} $B^0 - \bar B^0$ oscillations; 
(4) direct CP violation in semileptonic decays.

The existence of the first two classes of effects has been established; 
the third one most cleanly searched for through rate$(B^0 \bar B \to l^-l^-+X)$ $\neq$ 
rate$(\bar B^0 B \to l^+l^+ +X)$ is predicted to be very small and has not been observed yet 
\cite{KWON}. 
Within the SM there is no observable example of the forth class; even with a 
nondiagonal MNS matrix there is no realistic prospect of finding an asymmetry. Yet charged Higgs 
exchange could induce a transvers polarization for $\tau$ leptons in $B\to \tau \nu D^*$ or other 
T odd moments. 

{\bf (i)} $\phi_1$: The predicted CP  asymmetry in $B_d(t) \to [\bar cc] K_{S,L}$ has been firmly established 
and accurately measured: sin$2\phi_1 = 0.726 \pm 0.037$ \cite{KARIM}. It allows two solutions, namely 
\beqq
\phi_1^{CKM} = 23.28^o \pm 1.6^o   \; \;  \; \; \& \; \; \; \; \tilde \phi_1 = \pi/2 - \phi_1^{CKM}
\eeqq
The second solution $\tilde \phi_1$, which is clearly inconsistent with the allowed values for 
$|V(ub)|$ is disfavoured though not yet ruled out by the bounds on cos$2\phi_1$ extracted 
from $B_d(t)\to \psi K^*$. 

\noindent The most intriguing signal for the intervention of New Physics has surfaced in 
$B_d(t) \to \phi K_S$ 
(and similar channels driven by $b\to s q \bar q$ like $B_d(t) \to \eta K_S$) \cite{KWON}:   
KM theory predicts the same asymmetry as for $B_d(t) \to \psi K_S$, i.e. $C=0$, 
$S= {\rm sin}2\phi_1$, which is consistent with BABAR's findings -- 
$S=  0.50 \pm 0.25 ^{+0.07}_{-0.04}$, $C= 0.00 \pm 0.23 \pm 0.05$ -- yet in intriguing, though not 
conclusive discrepancy with BELLE's results of $S=0.06 \pm 0.33 \pm 0.09$, 
$C=-0.08 \pm 0.22 \pm 0.09$. I take this signal very seriously for the following reasons: 
(i) Since these transitions are driven by a loop process within the SM, they are natural candidates 
for revealing the intervention of New Physics. (ii) The SM prediction that $B_d(t) \to \phi K_S$ 
and $B_d(t) \to \psi K_S$ should exhibit a very similar CP phenomenology is reliable 
\cite{GROSS}. 
(iii) Other modes driven by the same quark-level transition operator $b\to s \bar s s$ exhibit 
similar `anomalies' \cite{KWON}. (iv) To make a sizeable deviation from the SM prediction for 
$B_d(t) \to \phi K_S$ compatible with the observed $\Gamma (B\to \gamma X)$ suggests the new 
effective transition operator to possess a different chirality structure, which would affect also 
the final state polarization in $B\to K^* \phi $, for which there are indeed 
intriguing indications in the data \cite{PIRJOL}. Of course future data will have to decide. 

{\bf (ii)} $\phi_3$: We are at a stage where the floodgates for many more CP asymmetries are 
opening \cite{KWON}. What we can 
say for sure right now is that {\em direct} CP violation has been observed in 
$\bar B_d/B_d \to K^{\mp}\pi^{\pm}$ by both BABAR and BELLE \cite{GRAZ}: 
\beqq 
\left. \frac{\Gamma (\bar B_d \to K^-\pi^+) - \Gamma (B_d \to K^+\pi^-)}
{\Gamma (\bar B_d \to K^-\pi^+) + \Gamma (B_d \to K^+\pi^-)}\right|_{BABAR\& BELLE} = 
-0.114 \pm 0.020
\label{DIRECTBDEC}
\eeqq
The promise of this mode and how the required final state interactions could be generated by 
the Penguin contribution was pointed out a long time ago in a seminal paper \cite{SONI}. The 
observed signal, Eq.(\ref{DIRECTBDEC}), is consistent with theoretical expectations; on general grounds it could not be much larger. Any value of $\phi_3$ 
extracted from it is highly model dependent -- at least until LQCD has been demonstrated to 
yield a reliable and accurate treatment of nonleptonic $B$ decays {\em including strong 
final state interactions}. 

\noindent One has to be aware of the constraints placed on direct CP violation due to 
CPT symmetry. For the latter imposes much more than equality of masses and lifetimes for 
particles and antiparticles, namely equality of rates for {\em sub}classes of CP conjugate 
channels, like the total semileptonic widths. Another relevant example is that 
$\Gamma (B^+ \to \pi^+ \pi^0) = \Gamma (B^- \to \pi^- \pi^0)$ has to hold up to isospin 
violating contributions (the final state is pure $I=2$). Furthermore the 
{\em weighted} CP asymmetries 
have to cancel between channels that can rescatter into each other \cite{CICERONE}. This 
relation is sometimes useful in practical terms, sometimes not.  For example 
$B \to K\pi \leftrightarrow D_s \bar D$; yet with the width for $B\to D_s \bar D$ so much 
larger than for $B\to K\pi$, a large CP asymmetry in the latter can be compensated for 
by a small one in the former. 

\noindent Great progress is being made in extracting $\phi_3$ from $B^{\pm} \to D^{neutral} K^{\pm}$. 
First pointed out as a mode exhibiting CP asymmetries driven by $\phi_3$ \cite{BSPAIS}, 
it was then reshaped into various versions of precision tools for extracting a value for 
$\phi_3$ \cite{ATWOOD}. The analysis has reached a new level now by involving the multibody final state $D^0/\bar D^0 \to K_S\pi^+\pi^-$ through a Dalitz plot   analysis  \cite{TIM}: 
\beqq 
\phi_3 = \left( 77 ^{+17}_{-19}|_{stat} \pm 13|_{syst} \pm 11|_{model}\right) ^o
\label{PHI3}
\eeqq

{\bf (iii)} $\phi_2$: We have a clear signal for a CP asymmetry in $B \to \pi^+\pi^-$ from BELLE. With \beqq 
\frac{R_+(\Delta t) - R_-(\Delta t)}{R_+(\Delta t) + R_-(\Delta t)} = S {\rm sin}(\Delta M_d \Delta t) - 
C {\rm cos}(\Delta M_d \Delta t) \; , \;  S^2 + C^2 \leq 1 
\eeqq
where $R_{+[-]}(\Delta t)$ denotes the rate for 
$B^{tag}(t)\bar B_d(t+\Delta)[\bar B^{tag}(t)B_d(t+\Delta)]$, they find 
\beqq 
S = - 1.00 \pm 0.21 \pm 0.07 \; , \; \; C = - 0.58 \pm 0.15 \pm 0.07 \; , 
\eeqq
which amounts to a 5.2 $\sigma$ CP asymmetry and {\em direct} CP violation 
with 3.3 $\sigma$ significance. For  {\em without} direct CP violation one would have to 
find $C=0$ and $S = - {\rm sin}2\phi_1$ \cite{ELE}. However this signal has so far not been confirmed 
by BABAR, which has found $S=-0.30 \pm 0.17 \pm 0.03$, $C=- 0.09 \pm 0.15 \pm 0.04$. Future data  will hopefully clarify this situation \cite{KWON}.

\section{Outlook}
\label{SUM}

It has become a seemingly tedious refrain at conferences that the SM has been very successful 
in describing data. Yet the successes it has scored recently in heavy flavour decays have a 
{\em qualitatively}  
new aspect: for they have provided the first decisive tests of the CKM description on a highly  
quantitative level. CKM parameters -- in particular $|V(cb)|$ -- have been determined with an 
accuracy seemingly unrealistic even five years ago; detailed error budgets have been provided 
by theorists; large CP asymmetries have been found where predicted. 

There are some intriguing deviations of data from expectations. Yet more important is that these novel 
successes of the SM do not resolve any of the central mysteries of the SM in particular in its 
mass generation for fermions -- they are `merely' built on them. Thus they strengthen the case for New 
Physics. 

There are actually two classes of New Physics, which are not necessarily related: 

\noindent $\bullet$ 
The New Dynamics driving the electroweak phase transition characterized by the 
${\cal O}$(1 TeV) scale. I find the theoretical case for its existence compelling and therefore 
refer to it -- in only slight hyperbole -- as the `guaranteed' New Physics ({\em g}NP). 

\noindent $\bullet$ 
The New Dynamics responsible for the flavour structure, the existence of families and their 
couplings. I find the case for its existence theoretically persuasive, yet we have no clear idea about its characteristic scales. Thus I name it the `strongly suspected' New 
Physics ({\em ss}NP). 

My friend Masiero likes to say: "You need luck to find New Physics." True, but one should 
remember Napoleon's quote: "Having luck (`fortune') is part of the job description for generals." 

The LHC and the Linear Collider are justified -- correctly -- to conduct campaigns for 
{\em g}NP. The latter is unlikely to shed light on {\em ss}NP, though it might. Yet the argument is reasonably turned around: detailed and comprehensive studies of flavour transitions can  
elucidate salient features of the {\em g}NP; the latter residing at the TeV scale could affect 
flavour dynamics significantly. One should keep in mind the following: one very popular example of 
{\em g}NP is supersymmetry; yet it represents an organizing principle much more than even a class 
of theories. I find it unlikely we can infer all required lessons by studying only flavour diagonal 
transitions. Heavy flavour decays provide a powerful and complementary probe of 
{\em g}NP. Their potential to reveal something about the {\em ss}NP is a welcome extra not required 
for justifying efforts in that direction. Therefore I see a Super-B and a Tau-Charm Factory as worthy 
siblings of the obvious daughter of the TEVATRON and LHC, the Linear Collider.

\section*{Acknowledgments}
I would like to thank the organizers of FPCP04 for creating such a fine meeting. 
This work was supported by the NSF under grant numbers PHY00-87419 \& PHY03-55098.


%
\label{IBigiEnd}

\end{document}